# SOCIAL PHYSICS: A MANIFESTO

J. Kertész [1], M. Barthelemy [2], G. Caldarelli [3], B.K. Chakrabarti [4], T. Di Matteo [5], S. Galam [6], M. Gonzalez [7], Y. Holovatch [8], J.A. Holyst [9], S. Jalan [10], H. Jeong [11], N. Johnson [12], R. Lambiotte [13], L. Lü [14], T. Lux [15], R.N. Mantegna [16], M. Perc [17], J. Saramäki [18], M. Small [19], M. Takayasu [20], T. Yasseri [21], W.-X. Zhou [22]

[1] DNDS, Central European University, Vienna, Austria, [2] IPhT, Université Paris-Saclay, France, CEA, CNRS, [3] Institute of Complex Systems, CNR-ISC, Rome, Italy, [4] CMPD, Saha Institute of Nuclear Physics, Kolkata, India, [5] Dept. of Math., King's College, London, UK, [6] Dept. for Political Research, Sciences Po, Paris, France, [7] Dept. of City and Regional Planning, University of Berkeley, CA, USA, [8] Yukhnovskii Institute for Condensed Matter Physics, National Academy of Sciences of Ukraine, Lviv, Ukraine, [9] Faculty of Physics, Warsaw University of Technology, Poland, [10] Dept. Physics, Indian Institute of Technology Indore, India, [11] Dept. Physics, KAIST, Daejeon, Korea, [12] Dept. Physics, George Washington University, Washington, DC, USA, [13] Math. Institute, University of Oxford, UK, [14] School of Cyber Sci. and Techn. of China, Hefei, China, [15] Dept. Economics, Christian Albrechts University of Kiel, Germany, [16] Dept. Physics and Chemistry, Universitá degli Studi di Palermo, Italy, [17] Faculty of Natural Sci. and Math., University of Maribor, Slovenia, [18] Dept. of Comp. Sci. Aalto University, Espoo, Finland, [19] School of Math and Statistics, University of Western Australia, Crawley, Australia, [20] School of Computing, Institute of Science Tokyo, Japan, [21] Centre for Sociology of Humans and Machines, Trinity College Dublin, Dublin, Ireland, [22] School of Business, East China University of Science and Technology, Shanghai, China

## Abstract

Social Physics seeks quantitative, empirically testable explanations of collective human behavior. Its name has a long and contested history, but its contemporary program is neither the claim that society is literally a physical system nor an attempt to replace the social sciences with physics. It is an interdisciplinary practice: observation and experimentation, model construction, mathematical and computational analysis, and repeated confrontation with data. The field has been transformed by analyzing digital traces, network science, agent-based modeling, large-scale experiments, and the capacity to study coupled dynamics across economic, social, technological, biological, and physical layers. We review advances in financial analysis and modeling; economic complexity; the structure and evolution of social networks; spreading, opinion, and information dynamics; inequality, cooperation, conflict, mobility, cities, culture, science, and collective behavior; and the emerging study of human – AI ecosystems. Across these domains, common features recur – heterogeneity, interaction, feedback, adaptation, non-equilibrium dynamics, multiscale organization, and emergent collective outcomes. We also confront the limitations that have arisen in this field: convenience samples, platform dependence, weak causal identification, over-universal claims, insufficient engagement with meaning and institutions, privacy risks, and the danger that predictive models become instruments of manipulation. We propose a future Social Physics organized around theory – data reciprocity; causal and generative explanation; measurement validity and cross-context generalization; responsible, participatory data governance; genuinely interdisciplinary collaboration; and reflexive study of interventions and algorithms that alter the systems being observed. Social Physics should aim not at immutable "laws of society," but at robust, scope-qualified regularities and mechanisms that enlarge human understanding and capacity for collective action.

## 1. Introduction: an old ambition – a new opportunity

Since Galileo, Huygens, and Newton helped establish what is now called "the method of physics", scientific disciplines have repeatedly aspired to its productive cycle: observation and experiment, abstraction and model construction, mathematical or computational analysis, prediction, and renewed confrontation with evidence. Its success has transformed both knowledge and society. Yet applying this method to human affairs has always appeared unusually difficult. Human beings interpret situations, anticipate one another, learn, organize, and change their behavior when classifications or predictions about them become known. The objects of inquiry are also subjects.

Newton's often-repeated lament after financial losses – that he could calculate the motions of heavenly bodies but not the madness of people – is probably apocryphal in its exact wording, but it captures a real contrast. Celestial mechanics offered stable objects, repeatable measurements, and powerful invariances. Markets, crowds, institutions, and cultures seemed contingent, reflexive, historically embedded, and resistant to compact law. Social Physics begins with the conviction that this contrast is not an excuse for abandoning quantitative explanation. Instead, it specifies the problem.

Already Hobbes sought a mechanistic account of political order [1]. The expression Social Physics emerged in the nineteenth century. Adolphe Quetelet, an astronomer and statistician, used it to describe the statistical regularities visible in births, deaths, crime, marriage, and other aggregate social phenomena [2]. His "average man" is the early appearance of the "representative agent" of today's mainstream economics, and it has been rightly criticized when treated as a norm rather than a statistical construct, yet his central empirical insight endures – individual unpredictability can coexist with population-level regularity. Auguste Comte had independently proposed "physique sociale" for a systematic science of society before coining the term sociology [3]. These projects mixed fertile methodological ambitions with the determinism and social prejudices of their times. Contemporary Social Physics must inherit ambition, however, without reductionism.

The exchange of ideas between physics and social inquiry has never been one-way. Nineteenth-century marginalist economics borrowed the language and mathematics of mechanics: utility resembled potential, scarcity acted like force, and equilibrium became a central organizing metaphor in the work of Walras and others [4]. Freud's early hydraulic model casts psychic life in terms of pressure, energy, and discharge. Richardson, a meteorologist, developed mathematical models of arms races and wars [5]. The concept of diffusion migrated into models of innovation and technology adoption, while the mathematical theory of diffusion later became fundamental to option pricing [6].

Conversely, social statistics influenced physics. Maxwell explicitly drew on regularities in large populations when constructing kinetic theory [7, 8]. The equilibrium concepts of Walras and Menger shaped the intellectual environment in which Gibbs developed his thermodynamic and statistical-mechanical ideas. Bachelier's mathematical theory of speculation anticipated key features of the random-walk treatment of Brownian motion [9]. Pareto's empirical work on income distribution tails helped make heavy-tailed distributions a recurring object across the sciences [10]. Social network analysis – from sociometry to the study of weak ties and block structure – provided concepts and empirical questions that profoundly shaped modern network science [11-13].

Ettore Majorana articulated an especially prescient position. In a paper published posthumously in 1942, he drew an analogy between the intrinsically probabilistic behavior of individual particles in quantum mechanics and that of humans and argued that statistical laws play a central role in both physics and social sciences [14]. Later, Herbert Simon's studies of bounded rationality and mechanisms generating skewed distributions [15, 16], and Benoit Mandelbrot's results on scaling and irregularity [17], prepared the ground for a science of complex social systems. No doubt, society is a complex system with many interacting constituents and emergent phenomena [18], but it has the additional difficulty for a scientific approach that humans are reflective, resulting in possible destruction of observed regularities

In the 70's and 80's, pioneering efforts were made to apply the physics approach to social phenomena. Prigogine and Herman published a book on the kinetic theory of traffic [19], Weidlich analyzed the interaction of social groups based on synergetics [20] and he became a promoter of modern Social Physics. Galam put forward general concepts on the overall role physics and physicists can play in investigating social problems and how that can influence physics itself [21], moreover, he coined the word "Sociophysics" [22] to distinguish it from the 19th century "Social Physics", however, recently the latter term is becoming increasingly used. It must be said that the interdisciplinary efforts of physicists were not widely recognized for a long time; in fact, from the very beginning until the late 1990s, the prospect of investigating human behavior from a physics-like perspective was rejected by most physicists [23, 24].

The modern expansion of Social Physics arose from a conjunction of developments. Statistical physics had matured a language and methodology for systems composed of many interacting elements; nonlinear dynamics and complexity science emphasized feedback, adaptation, and emergence; computers made large-scale simulation possible; network science supplied a representation of structured interactions; and digital technologies generated streams of behavioral data at previously inconceivable resolution, providing a historic opportunity [25-28]. Mobile phones, online platforms, administrative registers, remote sensing, and transactional records now allow researchers to study aspects of communication, mobility, consumption, cooperation, and collective attention across scales [29]. Massively multiplayer online games (MMOGs) provide a unique data source for studying virtual societies, recording all player actions and interactions at high temporal resolution over extended periods [30]. Traditional survey methods, as well as laboratory, field, and online experiments, are used to complement observation. Machine learning can discover structure, while mechanistic models can explain why it arises. In the 90's a separate branch of studies emerged focusing on finance and economics: econophysics. These developments resulted in a broad interdisciplinary research program, spanning the statistical characterization of economic and social systems, network-based approaches, and increasingly data-driven methodologies, while also highlighting important methodological and interdisciplinary challenges [31].

The word Physics is best understood methodologically, not territorially [32]. Sometimes the connection is direct: Ising spins inspire binary-choice and opinion models; percolation informs spreading and systemic risk; kinetic equations describe distributions; and phase transitions organize changes between collective regimes. But Social Physics is broader than transferring physical concepts and models. It is the perpetual interaction of measurement, experimentation, theory, and computation, with empirical evidence decisive.

Our thesis is that Social Physics is most useful when it seeks patterns, conditional regularities, causal mechanisms, and builds models from these data. Human systems exhibit constraints, conservation-like balances, scaling relations, robust distributions, tipping phenomena, and network effects – but also institutions, identities, power, norms, strategic interpretation, and technological mediation. These are not residual complications to be averaged away; they are constitutive causal structures.

This paper surveys the field's principal domains and sets out a manifesto for its next phase. The scope reflects that of the Journal of Social Physics [33]. The boundaries between sections are porous by design. Markets are social networks; epidemics reshape mobility and inequality; algorithms modify exposure and preferences; cities couple infrastructure, behavior, and institutions. The challenge is to understand these interdependencies without mistaking a large dataset for society itself. We do not claim to be comprehensive – our aim is to highlight representative achievements, identify methodological challenges, and outline future directions.

## 2. Finance, economics, and innovation

### 2.1 From stylized facts to market microstructure

Finance was the first major arena of modern econophysics because long, high-frequency, machine-readable price series were already available by the end of the last century. Early work decisively strengthened the non-Gaussian paradigm: returns show heavy tails, volatility clustering, long memory in absolute returns, and dependence across assets and time scales [34-37]. These "stylized facts" are not decorative deviations from an ideal model; they define the empirical target any plausible theory of markets must reproduce. Multiscaling [38, 39] and multifractal [40] analyses provide complementary frameworks for describing intermittent volatility across scales, although multifractality must be distinguished carefully from finite-sample effects, regime changes, and heavy-tailed marginals. More recent developments have moved beyond identifying multiscaling toward statistically robust estimation and testing of scaling behavior and its application to financial risk measures [41] while also investigating the relationship between price multiscaling and the rough-volatility paradigm [37].

The field moved from price statistics toward mechanisms. Empirical studies of price impact showed that trading affects prices in ways dependent on order size, execution strategy, liquidity, and the long memory of order flow [42]. Limit-order books revealed regularities in spread, depth, cancellation, placement, and execution times [43, 44]. Stochastic and empirical behavioral order-book models connect these microscopic events to mesoscopic price dynamics [45-47]. A detailed analysis of order-book data has shown that the fluctuation-dissipation relation holds for appropriately defined price fluctuations in close analogy with Brownian dynamics [48]. This progression – from a distributional regularity to a proposed generative mechanism and then to new measurements – is exemplary Social Physics.

Correlations among asset returns also provided a laboratory for network methods. Minimum spanning trees and related filtering procedures exposed hierarchical market organization and its evolution [49-51]. Higher order dependencies beyond pairwise correlations reveal hidden functional organization and community structures [52]. Random matrix theory offered null expectations for noisy covariance spectra and helped distinguish collective market modes and sectoral structure from sampling noise [53, 54]. Such tools have practical value, e.g., connecting correlation-based filtering with financial factor models [55], but their scientific importance is broader: they show how high-dimensional interdependence can be compressed without assuming that every observed association represents a stable relation.

Agent-based models (ABMs) reject the fiction of a representative, fully rational actor and model markets as adaptive populations. Threshold models, artificial stock markets, minority games, chartist–fundamentalist models, and minimal models have generated clustered volatility, fat tails, bubbles, and endogenous regime switching [56-61]. Their strongest contribution is not a claim that one specific artificial market reproduces reality. It demonstrates that macroscopic patterns observed in markets may emerge from interaction, heterogeneity, feedback, and institutional rules without an external shock of matching form. Applying the Langevin equation approach to the microscopic dynamics of ABMs theoretical predictions for the order book dynamics was calculated [62].

Economies need computational models that represent heterogeneity and institutions, especially outside equilibrium [63]. The relevant comparison is not ABM versus an idealized analytic theory in the abstract, but which model class answers specified questions with identifiable assumptions and validated consequences. ABMs nevertheless face demanding standards. Many parameter combinations can reproduce the same stylized facts; calibration may identify outcomes without identifying mechanisms; and additional behavioral detail can make a model less interpretable. The way forward combines micro-level evidence, disciplined calibration, out-of-sample validation, uncertainty quantification, model comparison, and intervention tests [64].

Economic systems are intrinsically networked: firms, banks, households, and countries interact through credit, trade, ownership, and supply chains. These connections determine how local decisions and shocks become aggregate outcomes, making network structure essential for understanding phenomena that cannot be captured by considering agents independently [65, 66].

## 2.2 Wealth, complexity, and systemic risk

Network theory is particularly important for systemic risk and resilience, where interconnections can both diversify risk and create channels for cascading failures. In financial systems, the probability of one institution's distress depends on its direct and indirect exposure to others, while incomplete knowledge of these connections can severely impair risk assessment. This makes network analysis not merely descriptive, but a necessary framework for connecting microeconomic interactions with macroeconomic stability [67].

Kinetic-exchange models demonstrated how distributions of money or wealth can emerge from simple exchange rules and heterogeneous saving propensities [68, 69]. Their analogy with gases is illuminating but incomplete: real wealth is produced, inherited, taxed, leveraged, institutionally protected, and embedded in unequal bargaining power. The challenge is to connect parsimonious distributional mechanisms to household panels, firm ownership structures, taxation, housing, education, and policy.

Economic complexity shifts attention from scalar wealth to productive capabilities. Networks linking countries, products, firms, occupations, or technologies reveal diversification, nestedness, and paths of structural transformation [70, 71]. Complexity measures can anticipate feasible diversification and illuminate why similar aggregate resources yield different development trajectories. Their interpretation, however, requires care: a ranking is not a causal theory, and trade records reveal capabilities only imperfectly.

Financial crises and supply disruptions underscore that risk is relational. DebtRank estimates distress amplification in interbank networks and formalizes the idea that institutions may be “too central to fail” [72]. Firm-level production networks allow systemic economic risk to be estimated from dependencies in the supply-chain [73]. These approaches replace individual robustness measures with system-level vulnerability. Future models must add multiplex exposures, substitution, inventories, contract duration, geography, expectations, and adaptive rewiring; stress tests should be evaluated against historical disruptions and used transparently when policy decisions affect firms and citizens. Models are expected to evolve into global tools that consider all aspects of the economy, including international trade, interactions between the financial sector and firms, and the role of innovations. The importance of working, verified, predictive models cannot be overestimated in anticipating and mitigating the effects of crises, and strengthening robustness of systems before shocks arrive.

### 2.3 Innovation and economic evolution

Innovation is a discovery process in a changing landscape. Empirical work on trade supports Schumpeterian creative destruction [74], while network approaches represent inventions, technologies, products, and knowledge domains as evolving relational systems [75]. The adjacent possible [76, 77] captures how existing combinations enable some novelties while excluding others; network models can formalize exploration, recombination, lock-in, and punctuated change. Yet innovation is not only search. Institutions allocate attention and capital, intellectual-property regimes change incentives, organizations restructure after discoveries, and cultural expectations shape what count as a valuable problem. Patent registers provide excellent data for studying regularities of the innovation process, including novelty profiles and future impact [78].

The field has progressed from fitting distributions to comparing mechanisms, from equilibrium models to adaptive non-equilibrium systems, and from isolated representative agents to heterogeneous multi-agent processes. The goal is to move from interpretation to forecasting and policy-relevant counterfactuals. Success will require more intense collaboration with economics. Analogies with other evolving adaptive systems will help establish a robust theoretical framework [79]. A better understanding of the complex system of innovation processes will lead to new, optimized strategies at the micro and macro levels.

## 3. The structure and evolution of social networks

### 3.1 Static networks and measuring relationships

Networks provide the natural language of interacting structures. Classic social science established that relations matter: weak ties connect communities and facilitate access to novel information [12]; small-world structure combines local clustering with short global paths [80]; and homophily organizes friendship, marriage, communication, and opportunity [81]. Network science expanded the scale and mathematical reach of this program, showing how degree heterogeneity [82], clustering, assortativity, communities, and centrality affect collective behavior [83, 84]. Studies have shown that social interactions span complex, weighted, signed, directed, temporal, multi-layer, and higher-order networks [85-89]. Complex network structures play an important role in several dynamical processes, including epidemic dynamics [90], innovation spreading [91], and political polarization [92].

Digital communication records transformed empirical work by enabling measurement of millions of time-stamped interactions [93]. Studies of mobile-phone networks revealed strong heterogeneity, community structure, and correlations between tie strength and network topology [94-96]. Observed calls, messages, follows, proximity events, and co-locations do not necessarily capture social ties of the same kind. Measurement models must specify how platform affordances, missing interactions, shared devices, bots, and sampling frames map observed traces onto social relations [97, 98]

### 3.2 Cognitive constraints and layered sociality

Humans do not have arbitrarily large personal networks. Evidence across offline and online settings supports cognitive and temporal constraints on active, socially relevant relationships, resulting in finite, possibly layered patterns of tie strength in egocentric networks associated with Dunbar's number [99-101]. The exact sizes and universality of these layers remain debated, but the broader finding is robust: brain capacity, attention, and time are scarce, relationships compete for maintenance, and the size of a network does not by itself determine the level of social support it provides.

This matters for diffusion, inequality, and well-being. A person may have many online relationships but few reliable strong ties; weak, bridging links provide access to diverse information while strong and cohesive relations provide trust and support. Kinship, household structure, workplace organization, migration, and life course. Longitudinal work should therefore model ego networks as resource-constrained, multiplex, and dynamically maintained systems [102].

### 3.3 Communities, overlap, and higher-order organization

Community detection [103] became a major area of research because mesoscopic organization mediates identity, exposure, coordination, and resilience. Modularity offers a powerful objective but also has a resolution limit [104] and other problems. Methods based on stochastic block models provide a more solid, statistically justifiable basis for community detection [105]. Link communities and overlapping methods recognize that individuals participate in multiple social contexts [106, 107, 104, 108]. Generally, no community structure should be considered as “the” solution of the detection problem. Communities can be structural, statistical, functional, perceived, or institutionally defined, and the method should match the question.

Many interactions are multiplex: the same pair may be kin, colleagues, friends, competitors, or communication partners. Others are irreducibly group-based: committee decisions, team production, group chats, or rituals cannot always be reduced to independent dyads. Multilayer networks and higher-order representations – hypergraphs and simplicial complexes – provide formalisms for these cases [109, 85]. While these structures can be handled within the realm of graphs [110], adopting an enriched framework may bring a new perspective for the modelling of processes and the design of algorithms. The context justifies special treatment, supported by added mathematical expressiveness and improved empirical evidence.

### 3.4 Time, causality, and network dynamics

Aggregating interactions into a static graph can create paths that never existed in time, erase burstiness, and misrepresent causality. Temporal network research shows that event order, circadian rhythms, bursty activity, tie turnover, and concurrency alter reachability and spread [86, 111]. Identifying relevant time scales is important for interpreting behavior and finding an optimal level of coarse graining [112]. Moreover, behavior and network structure coevolve: people influence one another, select similar partners, leave groups, and share contexts. Observational similarity cannot therefore be assigned automatically to social influence [113].

Disentangling influence, homophily, and shared environment remains a central methodological problem. Longitudinal models, natural experiments, randomized interventions, instrumental variables, and careful negative controls can help, but no single design is universally decisive. Network interference, i.e., the fact that the treatment of one node may affect the output of another one restricts the application of causal methods like Stable Unit Treatment Value Assumption (SUTVA). Social Physics can contribute by integrating explicit interaction models with modern causal inference rather than using prediction by high performing models as a substitute for causation.

## 4. Spreading, opinions, and collective attention

### 4.1 Simple and complex contagion

Random walks and diffusion are the simplest yet nontrivial dynamic processes on networks [114]. Epidemics of infectious diseases supply examples of similar spreading phenomena, with a long history of mathematical modeling [115]. Network topology, degree heterogeneity, modular structure, temporal ordering, and adaptive behavior all modify the spreading dynamics and thus the extent of the epidemic. Concepts from statistical physics, like phase transitions, percolation, and branching models, explain why connectivity can produce abrupt global reach as a function of a control parameter like transmission rate, while metapopulation and mobility models connect local contact patterns to spatial spread [115].

Mobility data, contact matrices, wastewater, case records, and genomic surveillance enabled unprecedented situational awareness during the COVID-19 pandemic, while behavioral response and policy altered transmission [116, 117]. Forecasts were conditional on future interventions and compliance, not prophecies. The lesson is that coupled disease-behavior dynamics are central: risk perception, trust, inequality, misinformation, and institutional capacity reshape contact networks and reporting. The interdisciplinary efforts, including contributions from physics, proved highly successful,as shown during the COVID-19 pandemic, when large-scale modeling informed government decision-making [118, 119].

In simple contagion, transmission is dyadic: transmission can occur through a single dyadic contact between an infectious and a susceptible individual. Complex contagion applies to behaviors, technologies, protests, and norms [120], where adoption may require multiple exposures or social reinforcement. Majority rules, thresholds, and repeated reinforcement characterize such processes. These two forms of spreading may occur simultaneously and distinguishing them requires appropriate empirical and modeling techniques [121].

### 4.2 Rumors, misinformation, and platform dynamics

Debates over vaccination, especially during the COVID-19 pandemic, provide an important setting for studying misinformation [122]and polarization [123]. Rumor models extend compartmental epidemiological models by adding forgetting, skepticism, and repeated contact [124]. Digital data have made it possible to trace cascades directly. Large-scale analysis found systematic differences in the diffusion of true and false news and highlighted the roles of novelty, emotion, and human sharing [125]. False and misleading information has become central to political discourse as new technologies accelerate their diffusion [126, 127]. Yet content labels, fact-checking, political context, and platform sampling can strongly shape results. Bots may amplify visibility [128], while recommendation and ranking determine the exposures from which users choose [129].

The key system is therefore not merely a user network but a coupled content–user–algorithm network. Generative AI deepens this coupling by turning platforms into hybrid human–AI networks, in which artificial agents serve simultaneously as transmission nodes and content producers [130]. Interventions can target producers, transmission channels, ranking rules, friction, media literacy, or incentives. Evaluations should measure not only immediate sharing but substitution, migration to other platforms, trust, and long-run adaptation.

### 4.3 Opinion dynamics, polarization, and echo chambers

Opinion dynamics translates mechanisms such as imitation, compromise, social pressure, bounded confidence, stubbornness, and repulsion into formal models. Many models, including voter, majority-rule, DeGroot, Friedkin-Johnsen, Deffuant, and Hegselmann-Krause models, [131-136] illuminate consensus, fragmentation, coexistence, and metastability. Opinions can be multidimensional, identity-laden, uncertain, strategically expressed, and anchored in material interests. The principal value of these models lies in isolating mechanisms rather than in treating expressed opinions as literal physical variables. It is usually very difficult to collect information about people's opinions – survey methods have their limitations in this field. Elections and the polls preceding them provide partial exceptions; in such settings, surprisingly simple models can sometimes have considerable explanatory power [137].

Opinion dynamics models are widely used to interpret political polarization, a pressing problem in many democracies [138]. Associated phenomena include mean separation of distributions, bimodality, or network segregation. Network topology can introduce unexpected effects, such as "more connections, higher polarization" [139].

Complex systems often exhibit characteristic precursors as they approach a tipping point including critical slowing down, increasing correlations, and loss of resilience. Such phenomena have been observed in social systems, like social media activity, collective sentiment, opinion polarization, or information cascades [140].

Digital technologies have vastly expanded communication opportunities. Yet they have not necessarily produced greater consensus and may, under some conditions, foster echo chambers and opinion bubbles. Echo chambers likewise combine selective exposure, homophily, reinforcement, and exclusion, and evidence varies across platforms and topics [141]. In adaptive homophilic networks, even positively biased recommendations may be unable to prevent fragmentation [142]. A robust Social Physics of polarization should jointly measure opinion, identity, network structure, content exposure, and institutional context.

### 4.4 Attention as a finite resource

News, memes, scientific papers, products, and political issues rise and decay as social reinforcement interacts with finite attention. When people are bombarded by a deluge of information, attention becomes a precious asset for which content providers fight. The result is a vicious circle that accelerates collective attention dynamics, with shorter periods of collective engagement with cultural items [143]. Another accompanying feature is the bursty nature of activities. Cascade size is heavy-tailed and difficult to predict from early dynamics alone. Models of priority queues, aging, preferential attachment, and novelty decay capture parts of this behavior [144]. The next step is to connect attention dynamics to information quality: which designs reward reliable information, support pluralism, and preserve space for slow deliberation rather than maximizing short-term engagement?

Attention is not allocated by an invisible hand. Recommendation algorithms now mediate exposure, ranking, and duration. They do not replace the dynamics above – priority queues, aging, preferential attachment, novelty decay – but operate on them, often accelerating burstiness and shortening attention spans by optimizing short-term engagement. Different algorithmic paradigms distribute attention differently, with consequences for diversity and fairness; the "diversity paradox" is discussed in Section 8.2 [129, 145]. Connecting attention to welfare therefore requires treating recommendation algorithms as constitutive mechanisms, not neutral filters. Yet designs reward reliable information and preserving slow deliberation is not merely technical: it depends on who defines welfare, how trade-offs are adjudicated, and whether governance resists capture. These questions sit at the intersection of Social Physics, Computational Social Science, and algorithmic governance—approaching finite attention from engineering implementation rather than dynamical mechanisms.

## 5. Social structure, inequality, cooperation, and conflict

### 5.1 Inequality as a networked process

Ending poverty is the first Sustainable Development Goal set by the United Nations. Targeted actions need accurate data; however, data about the most vulnerable populations is sparse and difficult to collect; data from new media may be useful but insufficient [146]. A combination of sources including digital communication data, satellite images, and online crowdsourcing can improve accuracy [147].

Inequality is multidimensional: income, wealth, health, education, status, exposure to hazards, political voice, and network access reinforce one another. Distributional summaries are essential but insufficient. Networks shape job information, credit, influence, collaboration, and mobility; spatial segregation and institutional sorting constrain whom people meet. Recently, this network aspect has proven crucial, as contact patterns that drive disease transmission are strongly influenced by people's socio-economic status [148].

Beyond developing advanced methods for data collection and analysis, Social Physics can contribute to multiscale models that connect micro-level interactions to macro-level patterns of stratification. Yet explaining how such patterns emerge is distinct from interpreting or legitimizing them. Models of inequality should therefore avoid naturalizing observed outcomes by presenting them simply as equilibria or power laws. Instead, they should make explicit the mechanisms and institutional arrangements that generate these outcomes, identify the groups differently affected by them, and, where possible, examine the consequences of feasible interventions. Similarly, aggregate distributional accuracy should be complemented by analyses of errors and effects across subgroups and interpreted in light of the relevant historical and institutional context.

Minorities and marginalized groups present a particularly important test [149]. Average behavior can conceal unequal exposure, network isolation, and asymmetric costs. Schelling's segregation model famously shows that mild local preferences can generate strong macro-segregation [150], but real segregation also reflects discrimination, zoning, wealth, violence, and policy. Minimal models reveal possibility; they do not allocate responsibility.

### 5.2 Cooperation and social dilemmas

Game theory was inspired by economics and has become a popular tool in the field, where the question of how rational individuals and institutions make strategic decisions under given conditions is of central importance [151]. Approaches from physics, focusing, e.g., on instabilities, phase transitions, and asymptotic stationarity, can help bridge models and observations [152].

Statistical physics has contributed substantially to evolutionary game theory and to the study of how cooperation persists despite incentives to defect [153]. Reputation, punishment, network reciprocity, group selection, and institutional design can all support cooperation [154, 155]. Structured populations create clusters of cooperators, but topology alone does not guarantee desirable outcomes. Payoff definitions, updating rules, information, mobility, interaction networks, and behavioral diversity matter.

Experiments increasingly connect formal games to actual behavior. Laboratory control identifies mechanisms, while field settings and large online experiments increase scale and improve ecological validity [156, 157]. Cross-cultural evidence warns against treating a convenient student sample as humanity. Future work should examine cooperation under inequality, identity, uncertainty, and human – AI interaction, where agents may delegate decisions or misperceive whether a partner is human.

### 5.3 Collective production, conflict, and violence

The importance of online collaboration is rapidly increasing. Platforms like Wikipedia or GitHub offer unusually detailed records of collective production. For example, Wikipedia edit histories reveal cooperation, reversion, norm formation, and "edit wars" [158]; open-source projects show how modular organization, reputation, and gendered behavior affect distributed work [159]. Such environments allow complete event histories within a platform, but not complete social histories: off-platform communication, institutional authority, and selection into participation remain hidden.

Conflicts more serious than "edit wars" range from interpersonal hostility to organized violence, terrorism, and wars. Continuing Richardson's quantitative tradition [5], Neil Johnson and collaborators identified dynamical patterns to estimate the escalation rate and timing of fatal attacks in insurgent and terrorist activity [160]. Modeling radicalization [161, 162] aims to find the basic mechanisms of a worldwide phenomenon and identify possible countermeasures. Contemporary work studies event data, insurgent organizations, protest diffusion, terrorist networks, and conflict forecasting. Heavy-tailed event size distributions and temporal clustering are common, but prediction has ethical and political consequences. False positives can stigmatize groups; models trained on reported events inherit surveillance and media biases. Work in this domain requires transparent uncertainty, independent oversight, and careful separation between explanation, early warning, and coercive deployment.

## 6. Mobility, cities, culture, and collective behavior

### 6.1 Human mobility and transportation

Mobile-phone, GPS, smart-card, and location-service data have complemented traditional survey information and revealed strong regularities in displacement, visitation frequency, and individual and collective mobility patterns [163-165]. Radiation and gravity models connect flows to population and opportunity distributions [166]. These findings are relevant for epidemic modeling, transportation planning [167], and urban science. It should be emphasized that these data sources were not generally designed to analyze mobility; thus, their use is indirect – for example, in determining the workplace and home – which may result in inaccuracies. Further methodological development, purpose-built data, and careful data integration are needed to improve predictions, a process in which important but not self-evident.

Mobility is constrained by infrastructure, care responsibilities, disability, income, borders, and safety. A mature account therefore moves beyond universal trajectories to heterogeneous capability. Explanatory models should represent trip purpose, accessibility, schedule, household coordination, and policy. Privacy-preserving aggregation, synthetic data, secure research environments, and data minimization are prerequisites, not afterthoughts.

### 6.2 Cities as coupled social – physical systems

Cities are complex, evolving systems that share some features with developing organisms [168, 169]. Quantitative approaches to cities seek to identify regularities in urban structure and function using statistical analysis, spatial data, network methods, and mathematical models. These approaches have revealed systematic patterns in infrastructure, mobility, economic activity, innovation, and other urban indicators. Urban scaling is one such framework, relating these quantities to city population in a compact comparative form [170]. More generally, however, quantitative regularities depend on how cities, observables, and spatial and temporal scales are defined. Statistical relationships, including scaling exponents, summarize empirical outcomes rather than explain them, and theoretical models are therefore needed to identify the mechanisms that generate the observed patterns [171].

Polycentricity, land use, transport, housing markets, amenities, segregation, and governance jointly produce urban outcomes. The interplay of these factors results in surprising effects, e.g., how urban geography and human-made structures influence socioeconomic inequalities [172].

Cities are exemplary multilayer systems [173]. Transport disruption changes access to employment; land prices alter residential sorting; amenities influence mobility; social networks mediate information and support; heat and pollution produce unequal health burdens. Digital twins and high-resolution models are promising only if they represent uncertainty, disadvantaged populations, and institutional decisions – and if affected communities can contest their assumptions [174]. In such models, the coupling between social and physical processes is increasingly mediated by a computational layer that measures, models, and modifies how they interact [175]. This indirect coupling closes feedback loops, making urban governance a problem of system-wide dynamics rather than layer-by-layer optimization.

Furthermore, replicating a city as a digital twin does not necessarily mean understanding how the city works. Even a highly detailed representation may reproduce its components and dynamics without revealing the mechanisms that generate collective behavior. Cities are complex systems in which mobility, economic activity, infrastructure, and social interactions continuously influence one another across multiple scales. [176] Understanding therefore requires moving beyond replication to identify the interactions, feedback loops, and network structures responsible for emergent phenomena. In the best case, a digital twin tells us *what is happening*; a theory of the underlying networks is needed to explain *why it happens* and, ultimately, to predict how the city will respond to change [177].

### 6.3 Traffic, crowds, and panic

Traffic flow, pedestrian motion, and evacuation are domains where physical constraints and social decisions meet directly. Physics-based theories like fluid-like descriptions [178] or car-following models [179] of vehicular traffic flow on highways have a long history, and together with the later cellular automata approach [180] they have the goal to reproduce and explain jams, lanes, stop-and-go waves, and collective alignment [181-183]. Urban traffic presents a different challenge with interesting network effects [184], multilayer properties [185], and relations between jamming and critical phenomena [186]. Railway networks provide another domain for complexity research, from delay minimization [187] to the study of passenger choices [188]. Social-force models of pedestrian motion offered influential accounts of bottlenecks and escape dynamics [189]. All these topics are closely related to direct applications and optimization tasks in urban planning and traffic management.

Under high-density conditions, the movement of dense human crowds can closely resemble granular flow, despite the intentional nature of individual movement. In emergencies, collective behavior can be understood by assuming simple pairwise "forces" between people [189]. As crowd density increases, collective motion may become more constrained and particle-like. Such models have already been used successfully in engineering practice to plan emergency routes in high-occupancy buildings and venues such as theatres or stadiums; however, the potential of this physics-based approach is far from fully exploited. Another physics-based reasoning is to consider people's movements in cities as electric current, which helps explain spatiotemporal patterns of human mobility [190].

### 6.4 Language, culture, and diversity

Since Zipf's pioneering work, researchers have intensively studied statistical regularities, often in the form of scaling laws, in languages [191]. Languages evolve through interaction, demography, migration, status, education, and media. Competition models can explain extinction thresholds or coexistence, while network and spatial models represent contact and bilingualism [192]. Cultural evolution similarly examines how traits vary, transmit, recombine, and are selected, as often studied within the framework of the Axelrod model; see [193]. These models are strongest when cultural traits are not treated as interchangeable particles: meaning, prestige, identity, and power influence transmission.

Research has also identified regularities in cultural products. Network analysis has studied the social structure of novels and compared it to real-world society [194, 195]. Researchers have statistically investigated temporal patterns in different genres of music [196, 197] and found similarities and differences, all useful for artificial music production. Statistical characterization of visual art [198] uncovered significant differences between classic and contemporary paintings. Although works of art usually result from individual creativity, their reception is a social phenomenon. The distinction between value and success is crucial in this context where the latter is easier to measure. The path to success is complex, involving numerous factors ranging from individual abilities to the role institutions play [199].

Global connectivity can homogenize some practices while increasing accessible diversity and enabling new niches. Diversity should be measured at multiple levels – individual repertoires, local communities, platforms, and global production. Recommendation systems now act as cultural intermediaries, making the study of cultural dynamics inseparable from algorithms.

### 6.5 Aging, health, sport, and animal societies

Social Physics also encompasses aging [200] and circadian behavior [201], health networks [202], sports performance and rankings [203, 204], and animal collective behavior like flocking [205]. The criticality hypothesis posits that animal collectives—such as fish schools, bird flocks, and insect swarms—operate near a (pseudo-)critical point, thereby optimizing their capacity for collective computation and information processing [206]. Wearables provide longitudinal signals of activity and sleep; sports offer clearly defined outcomes and repeated competition; animal groups enable controlled observation of coordination. Ranking systems based on pairwise competition or prestige are useful but can create incentives and feedback. Across these topics, the common opportunity is to connect multiscale measurements to interpretable mechanisms while resisting the temptation to equate what is readily measured with what matters.

## 7. Production and organization of knowledge

### 7.1 Science of Science

Science is itself a social system of discovery, evaluation, collaboration, competition, and cumulative advantage—and therefore an appropriate subject of research [207, 208]. Citation and collaboration networks reveal team formation, disciplinary boundaries, mobility, novelty, and the diffusion of ideas [209, 210]. Preferential attachment helps explain skewed citation distributions; team science has expanded; and atypical combinations of knowledge can be associated with high impact [211-213]. Models of career trajectories and "hot streaks" seek regularities in individual success [214]. The decomposition of success measures to skill-related and random factors has been generalized beyond science [215, 199].

These analyses can illuminate institutions, but they can also intensify metric fixation. Citations are field- and time-dependent, authorship conventions differ [216], databases have coverage biases, and impact does not equal quality. The use of scientometric indicators has become popular for evaluating journals, institutions, and even scientists, raising the question of how Goodhart's law [217] influences the development of science in general and individuals' careers in particular [218], especially when rankings determine funding or employment. A responsible science of science studies this reflexivity and evaluates reforms – lotteries, narrative CVs, double-anonymous review, open peer review, team credit, and resource redistribution – rather than merely optimizing predictive rankings.

### 7.2 Knowledge space: organization and navigation

Human development is based on accumulating knowledge, which is stored and made searchable, e.g., in libraries and encyclopedias. The body of the complex system of human knowledge grows rapidly; specialized techniques are needed to organize, uncover [219, 220], and visualize [221 the structure and organization of the knowledge space. Dynamic network models and processes on them provide helpful tools to approach this Digital technologies and the web have dramatically increased storage and search capacity; understanding the laws of orientation in the knowledge space is pivotal. Wikipedia, the free encyclopedia, is the world's largest collection of information, and it records all changes and discussions on cooperatively edited articles; thus, it provides an ideal medium for studying regularities in the production, access, search, and navigation of information [222, 223]. Wikipedia games [224] are helpful in this respect, where players solve the problem of finding page B from page A through Wikipedia hyperlinks. Researchers have examined the origins of popularity, quality assessment, editing habits, geographic differences, and how navigation strategies depend on personality in this extraordinary knowledge collection [158].

## 8. Human-AI ecosystems: the new frontier

### 8.1 Impact of algorithms on society

Artificial intelligence is changing the ontology of Social Physics [225]. Algorithms are not only instruments used by researchers or passive channels between humans. They select information, recommend partners and products, set prices, generate cultural content, converse, negotiate, and increasingly act as participants in social systems. The relevant unit is a human–AI ecosystem in which users, institutions, data, models, interfaces, and rules coevolve [226].

Automated selection of content and partners in the new media causes algorithmic bias, leading to echo chambers and enhanced polarization [227]. Bounded-confidence and coevolving-network models provide useful starting points [228], but empirical identification requires platform audits, controlled experiments, and data access. The scientific goal is not just to prove that "algorithms cause polarization" in general, but to map the conditions under which specific objectives, architectures, and user responses amplify or damp particular dimensions of polarization. Optimizing click probability in a static test set cannot answer whether a system narrows political or cultural diversity, amplifies inequality, facilitates discovery, or changes identity over years.

### 8.2 Human–AI coevolution

Recommender-system research has long framed recommendations as an information-filtering problem and developed collaborative, content-based, diffusion, and network approaches [129]. Social Physics adds a dynamical question: how does recommendation alter the preferences, networks, inventories, and behavioral data on which subsequent recommendations depend? A remarkable effect is the "diversity paradox": While the use of recommendation systems increases individual diversity of choices, the global diversity of choices in the population shrinks [145].

Navigation systems change routes and congestion; marketplaces change seller strategy and product visibility; scoring systems change who applies and how institutions allocate opportunity; generative assistants shape language and knowledge production. These are adaptive interventions at scale, often without an untreated control group. Algorithmic curation creates feedback. A recommender trained on past engagement changes future exposure; changed exposure changes behavior; new behavior becomes training data. Human–AI coevolution formalizes this feedback-loop perspective [226]. In extreme cases, these feedback loops can cause generative AI models to collapse [229]. New ideas from physics and complexity science may show how to mitigate the related negative effects.

### 8.3 Machine culture and hybrid collective intelligence

Machine culture denotes cultural traits mediated or generated by machines and the transformation of variation, transmission, and selection by intelligent systems [230]. Generative models learn from human cultural archives, produce new material, and feed future training corpora. This creates loops of imitation, homogenization, novelty, and possible model collapse. Provenance, intellectual property, linguistic diversity, and the visibility of small cultures become system-level questions.

AI-enhanced collective intelligence asks when hybrid groups outperform humans or machines alone [231]. Complementarity is possible: humans contribute contextual judgment, values, diverse experience, and responsibility; AI contributes search, memory, simulation, and scalable coordination [232]. But adding AI can also anchor groups, centralize failure, suppress minority information, or create automation bias. Performance must be evaluated not just by average accuracy but by robustness, diversity, calibration, legitimacy, and the distribution of benefits and errors.

Recent work proposes a Social Physics explicitly oriented toward advanced AI, including cooperation and norms in mixed populations, language–behavior coevolution, delegation, and governance [233, 234]. This agenda should include gender and other structural inequalities. AI systems inherit imbalances from data, occupational segregation, evaluation practices, and product design; their adoption can redistribute paid and unpaid work. Bias is not only a property of model outputs but an emergent feature of deployment networks.

### 8.4 Experiments and models of human–AI cooperation

AI raises justified societal concerns. The capabilities of generative models and their rapid development challenge existing forms of human oversight. Understanding should precede intervention, and Social Physics can contribute by identifying feedback mechanisms, collective risks, and effective governance options.

Controlled experiments are needed to show how people respond to advice, disclosure, anthropomorphism [235], and uncertainty [236]. Longitudinal platform studies can measure preference and network change, ideally through privacy-preserving researcher access [237]. Agent-based and game-theoretic models can explore feedback, strategic adaptation, and rare systemic outcomes [238].

Generative agents extend this capacity by allowing theories themselves to be iteratively simulated and revised against data [239]. The field should resist two symmetric errors: treating AI as an autonomous species divorced from owners and infrastructures or treating it as a neutral tool whose effects reduce to user intent. AI agents have designed objectives, trained dispositions, resource costs, institutional principles, and affordances that constrain action. Social Physics can make these relations explicit and ask how alternative designs change collective regimes.

### 8.5 Synthetic populations

Generative models can produce synthetic personae with selected real-world characteristics and simulate their interactions as artificial populations [240]. Interesting applications include persona-based simulations in social science, economics, and marketing research, as well as exploring business and policy decisions. Recent research focuses on pretraining details [241], scalability [242], and the emergence of social rules in LLM-generated societies [243, 244], as well as integrated paradigms that cycle between theory construction, generative simulation, and empirical validation [239]. LLMs playing social games against each other provide a simplified but instructive setting to study similarities and differences between human and artificial populations [245].

The central question is whether LLM outputs track real population distributions or just plausible-sounding text. Although persona-conditioned prompting can produce response distributions resembling those of specific demographic subpopulations, substantial caveats remain – LLMs perform better when explicit features drive human judgments but may diverge when faced with competing intuitions and generated data diversity remains a persistent challenge because LLMs tend to produce homogeneous outputs [246]. Closely related issues are bias propagation and fairness. At a more fundamental level, questions like calibration, controllability, and reproducibility arise. The epistemological grounding of LLM-based synthetic data is also questioned, as it rests on different assumptions than traditional synthetic data methods.

## 9. Methodological commitments for the next Social Physics

Social Physics has achieved remarkable results by combining physics-inspired measurement and modeling. Its record also contains cautionary episodes: search and social-media signals have overpredicted real-world trends [247]; platform populations have been mistaken for national populations [248]; social influence has been inferred from correlated behavior [249]; research centered on the United States remains overrepresented [250]; and physicists have occasionally rediscovered phenomena already familiar to social scientists [251, 252]. These limitations motivate the following commitments.

### 9.1 Begin with valid measurement and make heterogeneity central

Every digital trace is generated by a sociotechnical process. A “friend,” “share,” “view,” “location,” or “sentiment” is an operationalization shaped by an interface, business model, and data pipeline. Construct validity requires engagement with surveys, ethnography, domain knowledge, and qualitative interpretation. Triangulation can reveal whether a convenient proxy tracks the intended phenomenon.

Large sample size reduces sampling variance but does not cure selection bias, measurement error, confounding, or distribution shift.

Individuals differ in resources, identities, preferences, constraints, and exposure; places and institutions also differ. Distributional outcomes and subgroup effects should therefore accompany averages. Heterogeneity is not merely noise around the "real" mean – it may be the central result.

### 9.2 Close the loop between theory, data, explanation, and causality

Data should challenge models, and models should determine what new data is needed. Purely deductive toy models and purely inductive pattern mining are both incomplete. The preferred cycle is: empirical regularity, candidate mechanism, discriminating prediction, new measurement, and revision.

Prediction and explanation are complementary but distinct. A flexible model may forecast without identifying a transportable mechanism; a simple generative model may clarify a mechanism while predicting poorly at the individual level. Claims should state their target – description, forecast, causal effect, mechanism, counterfactual, or policy optimization – and causal claims must be commensurate with the research design. Experiments, quasi-experiments, causal inference under interference, and explicit generative models should be combined where possible.

### 9.3 State scope conditions and embrace pluralism

Social regularities vary across institutions, cultures, periods, and scales. Rather than announcing universality from one platform or population, researchers should specify scope conditions and pursue cross-context replication. Model ensembles and competing explanations are often more informative than a single universal curve.

### 9.4 Treat networks as measured, dynamic systems and study reflexivity

Relations have type, strength, direction, duration, context, and uncertainty. Networks coevolve with behavior and institutions; multilayer and higher-order models should be used when they represent measured mechanisms, not merely because formalism is fashionable. Predictions, rankings, and recommendations also alter behavior. Models deployed in credit, employment, policing, media, and science become components of the systems they describe, making feedback, strategic response, and performance primary research objects.

### 9.5 Make ethics, privacy, and power part of the research design

Public availability does not imply ethical availability. Reidentification, group harm, surveillance, and downstream misuse remain possible after names are removed. Ethical review should cover collection, linkage, inference, sharing, retention, publication, and deployment. Data minimization, secure access, appropriate privacy-enhancing technologies, and participatory governance can reduce risk. Researchers must also ask who can observe whom: exclusive corporate data may create a science in which powerful institutions see populations while populations cannot inspect institutions.

The field needs durable public-interest infrastructures that enable independent, privacy-preserving research on socially consequential platforms and administrative systems. Access should not depend solely on corporate discretion or personal connections; governance should include researchers, data holders, regulators, and affected publics.

### 9.6 Require reproducibility, robustness, and calibrated uncertainty

Open data and code are important but may conflict with privacy or contractual constraints. Reproducibility then requires layered strategies: executable code, synthetic or strongly aggregated data, preregistration, detailed provenance, secure enclaves, independent replication, and robustness tests. Because platforms change interfaces and policies, temporal reproducibility and versioned measurement are essential.

Uncertainty should be propagated and communicated. Scenario ranges, stress tests, sensitivity analyses, and qualitative failure modes often serve decision-makers better than a single precise forecast. Rare events, structural breaks, and model failure deserve explicit treatment.

### 9.7 Build genuinely interdisciplinary cooperation

Interdisciplinarity means shared problem formulation, not borrowing terminology after the fact. Physicists contribute abstraction, scaling, dynamics, and generative modeling; social scientists contribute theories of institutions, meaning, measurement, causality, and power; computer scientists contribute algorithms, data systems, and formal verification; domain experts and communities contribute indispensable situated knowledge. Collaboration should begin when the question and measurement strategy are defined, not only after the model has been built.

### 9.8 Orient the field toward collective capability

The ultimate measure of Social Physics is not its ability to predict clicks or rank people. It is whether it helps societies understand interdependence, anticipate systemic risk, design fairer institutions, coordinate under uncertainty, and preserve human agency. Applications should be evaluated by public value, legitimacy, and the distribution of benefits and harms.

## 10. Future directions

These commitments open several research frontiers. The following list is necessarily incomplete.

*Causal dynamics across scales.* We need formalisms linking cognitive and relational mechanisms to organizations, cities, markets, and institutions, while allowing feedback between levels. Special care is needed if renormalization or coarse graining is applied to social systems, because social aggregation is rarely lossless and must preserve meaning and agency of entities.

*Coupled crises.* Pandemics, climate hazards, migration, supply disruptions, financial stress, misinformation, and political conflict interact. Multilayer models should study compound risk, adaptation, and unequal vulnerability, supported by common scenario protocols and retrospective validation.

*AI for tipping point in social dynamics:* AI models need to be refined to predict real-world critical transitions in noisy, real-time data while addressing the reflexivity problem, where predictions inherently alter human behavior.

*Adaptive interconnected systems.* Financial markets, institutions, firms, and production networks are increasingly interconnected across multiple scales and layers. Political systems depend on economic players, social movements, climate changes. A key challenge is to develop data-driven models that capture the coevolution of dependencies, identify emerging systemic vulnerabilities, and distinguish transient correlations from persistent structural relationships.

*Experiments in networks.* Designs that recognize spillovers, changing ties, heterogeneous treatment, and ethical constraints will be central. Digital experiments should be paired with field knowledge and long-term outcome measurement.

*Foundation models as scientific instruments and social actors.* Large language and multimodal models can extract information, simulate agents, and support theory generation, but synthetic agents are not validated representations of humans. Benchmarking should compare simulated and observed behavior across contexts, detect training-data leakage, and quantify where models reproduce stereotypes or collapse diversity.

*Function of institutions*. Rules, organizations, law, and governance shape interaction topologies and incentives. Social Physics has often modeled agents on a fixed network; the next generation must model how institutions create networks, define categories, enforce decisions, and become contested.

*Climate and sustainability transitions.* Technology adoption, collective action, mobility, production networks, and public opinion determine whether decarbonization succeeds. Models must integrate material flows with social diffusion and distributive politics.

*Comparative and global Social Physics.* More work must be conducted with and for populations beyond wealthy, digitally saturated societies. Research partnerships should build local capacity, respect data sovereignty, and avoid extractive collection. Cross-cultural variation can reveal mechanisms invisible in a single setting.

*Theory of responsible intervention.* Many systems admit control in a mathematical sense, but legitimate social steering is not equivalent to controllability. Future work must unite optimal intervention with rights, consent, transparency, contestability, and democratic oversight.

*Dynamics of human–AI ecosystems.* We need new methods to continually measure the interactions between humans and AI (e.g., recommender systems) as well as new mathematical models and theoretical tools to describe the coevolution of human–AI systems.

*Establishing ethical framework for human-AI interaction.* Interdisciplinary collaboration, quantitative modelling tools, and counterfactual reasoning should produce policy recommendations for a safer and more human-friendly world.

## 11. Conclusion

Social Physics has advanced because it recognized that aggregate social regularities can emerge from interaction among heterogeneous agents, and because new data and computation made those interactions increasingly observable. It has contributed to powerful descriptions of markets, networks, diffusion, mobility, cities, cooperation, conflict, and science.

The field now faces a decisive transition. Data are abundant but unevenly owned; models are powerful but can be opaque; algorithms are becoming participants in the systems under study; and prediction can change behavior. Under these conditions, methodological sophistication without reflexivity is not enough. We therefore understand Social Physics not as physics applied to passive social matter, but as a collaborative science of interdependent, adaptive, meaning-making systems. Its future will be secured not by claiming exclusive jurisdiction over social phenomena, but by building a common empirical and theoretical language across disciplines.

The Journal of Social Physics is intended as a forum for that program: open, plural, quantitatively rigorous, empirically accountable, and engaged with the consequences of knowledge. This manifesto is an invitation – to test its propositions, extend its domains, expose its blind spots, and help construct a Social Physics worthy of the societies it seeks to understand.

**Author Contributions:** J. Kertész: Concept, writing the first draft, editing, & finalizing; M. Barthelemy: Writing - review & editing; G. Caldarelli: Writing - review & editing; B.K. Chakrabarti: Writing – review & editing; T. Di Matteo: Writing - review & editing; S. Galam: Writing - review & editing; M. Gonzales: Writing - review & editing; Y. Holovatch: Writing - review & editing; S. Jalan: Writing - review & editing; H. Jeong: Writing - review & editing; R. Lambiotte: Writing - review & editing; L. Lü: Writing - review & editing; J. Saramäki: Writing - review & editing; M. Takayasu: Writing - review & editing; T. Yasseri: Writing - review & editing; W.-X. Zhou: Writing - review & editing; N. Johnson: Review & editing; T. Lux: Review & editing; R.N. Mantegna: Review & editing; M. Perc: Review & editing; M. Small: Review & editing

**Funding:** JK: ERC grant No. 810115-DYNASNET; MP: ARIS grant No. P1-0403;

**Institutional Review Board Statement:** Not applicable

**Informed Consent Statement:** Not applicable

**Data Availability Statement:** Not applicable

**Acknowledgement: –**

**Conflicts of Interest:** The authors of this paper are Board Members of J. Soc. Phys. They had no access to information regarding its peer-review process. The paper was handled by a Guest Editor specifically appointed for this process.

**Use of AI and AI-Assisted Technologies:** LLM was used to assist with source citation searches, for linguistic editing, and final formatting.

**References**

[1] Hobbes, T. De Cive. Paris, 1642.

[2] Quetelet, A. A Treatise on Man and the Development of His Faculties. William and Robert Chambers, Edinburgh, 1842.

[3] Comte, A. Cours de philosophie positive, Vols. 1-6. Bachelier, Paris, 1830-1842.

[4] Mirowski, P. More Heat than Light: Economics as Social Physics, Physics as Nature's Economics. Cambridge University Press, Cambridge, 1989.

[5] Richardson, L.F. Arms and Insecurity: A Mathematical Study of the Causes and Origins of War. Boxwood Press, Pittsburgh, 1960.

[6] Black, F.; Scholes, M. The pricing of options and corporate liabilities. J. Polit. Econ. 1973, 81, 637-654.

[7] Porter, T.M. The Rise of Statistical Thinking, 1820-1900. Princeton University Press, Princeton, 1986.

[8] Maxwell, J.C. Molecules. Nature 1873, 8, 437-441.

[9] Bachelier, L. Théorie de la spéculation. Ann. Sci. Éc. Norm. Supér. 1900, 17, 21-86.

[10] Pareto, V. Cours d'économie politique, Vol. 2. F. Rouge, Lausanne, 1897.

[11] Moreno, J.L. Who Shall Survive? A New Approach to the Problem of Human Interrelations. Nervous and Mental Disease Publishing, Washington, DC, 1934.

[12] Granovetter, M.S. The strength of weak ties. Am. J. Sociol. 1973, 78, 1360-1380.

[13] Wasserman, S.; Faust, K. Social Network Analysis: Methods and Applications. Cambridge University Press, Cambridge, 1994.

[14] Majorana, E. The value of statistical laws in physics and social sciences. Quant. Finance 2005, 5, 133-140. (English translation by R.N. Mantegna of the 1942 Scientia article.)

[15] Simon, H.A. A behavioral model of rational choice. Q. J. Econ. 1955, 69, 99-118.

[16] Simon, H.A. On a class of skew distribution functions. Biometrika 1955, 42, 425-440.

[17] Mandelbrot, B.B. The variation of certain speculative prices. J. Bus. 1963, 36, 394-419.

[18] Ball, P. Why Society Is a Complex Matter. Springer, Berlin, 2014.

[19] Prigogine, I.; Herman, R. Kinetic Theory of Vehicular Traffic. Elsevier, New York, 1971.

[20] Weidlich, W. Dynamics of interacting social groups. In Synergetics: Cooperative Phenomena in Multi-Component Systems; Haken, H., Ed.; Springer, Berlin, 1974.

[21] Galam, S. Physicists as a revolutionary catalyst. Fundam. Sci. 1980, 1, 351-353.

[22] Galam, S.; Gefen, Y.; Shapir, Y. Sociophysics: A new approach of sociological collective behavior. I. Mean-behavior description of a strike. J. Math. Sociol. 1982, 9, 1-13.

[23] Galam, S. Sociophysics: A personal testimony. Physica A 2004, 336, 49-55.

[24] Galam, S. Physicists, non-physical topics, and interdisciplinarity. Front. Phys. 2022, 10, 986782.

[25] Galam, S. Sociophysics. Springer, Berlin, 2012.

[26] Pentland, A. Social Physics. Penguin Books, New York, 2014.

[27] Sen, P.; Chakrabarti, B.K. Sociophysics: An Introduction. Oxford University Press, Oxford, 2014.

[28] Jusup, M.; Holme, P.; Kanazawa, K.; et al. Social physics. Phys. Rep. 2022, 948, 1-148.

[29] Lazer, D.; Pentland, A.; Adamic, L.; et al. Computational social science. Science 2009, 323, 721-723.

[30] Szell, M.; Thurner, S. Measuring social dynamics in a massive multiplayer online game.Soc. Netw. 2010, 32, 313-329.

[31] Kutner, R.; Ausloos, M.; Grech, D.; et al. Econophysics and sociophysics: Their milestones and challenges. Physica A 2019, 516, 240-253.

[32] Holovatch, Y.; Kenna, R.; Thurner, S. Complex systems: Physics beyond physics. Eur. J. Phys. 2017, 38, 023002.

[33] Journal of Social Physics. Aims and Scope. Available online: https://www.sciltp.com/journals/jsp/aimscope (accessed September 30, 2026).

[34] Mantegna, R.N.; Stanley, H.E. An Introduction to Econophysics: Correlations and Complexity in Finance. Cambridge University Press, Cambridge, 2000.

[35] Cont, R. Empirical properties of asset returns: stylized facts and statistical issues. Quant. Finance 2001, 1, 223-236.

[36] Lux, T. Applications of statistical physics in finance and economics. In Handbook of Research on Complexity; Rosser, J.B., Jr., Ed.; Edward Elgar, Cheltenham, 2009; pp. 213-258.

[37] Johnson, N.F.; Jefferies, P.; Hui, P.M. Financial Market Complexity. Oxford University Press, Oxford, 2003.

[38] Di Matteo, T. Multi-scaling in finance. Quant. Finance 2007, 7, 21-36.

[39] Brandi, G.; Di Matteo, T. Multiscaling and rough volatility: An empirical investigation.Int. Rev. Financ. Anal. 2022, 84, 102324.

[40] Jiang, Z.-Q.; Xie, W.-J.; Zhou, W.-X.; et al. Multifractal analysis of financial markets: a review. Rep. Prog. Phys. 2019, 82, 125901.

[41] Brandi, G.; Di Matteo, T. On the statistics of scaling exponents and the multiscaling value at risk. Eur. J. Financ. 2022, 28, 1361-1382.

[42] Bouchaud, J.-P.; Farmer, J.D.; Lillo, F. How markets slowly digest changes in supply and demand. In Handbook of Financial Markets: Dynamics and Evolution; Hens, T.; Schenk-Hoppé, K.R., Eds.; North-Holland, Amsterdam, 2009; pp. 57-160.

[43] Bouchaud, J.-P.; Mézard, M.; Potters, M. Statistical properties of stock order books: empirical results and models. Quant. Finance 2002, 2, 251-256.

[44] Eisler, Z.; Kertész, J.; Lillo, F.; et al. Diffusive behavior and the modeling of characteristic times in limit order executions. Quant. Finance 2009, 9, 547-563.

[45] Mike, S.; Farmer, J.D. An empirical behavioral model of liquidity and volatility. J. Econ. Dyn. Control 2008, 32, 200-234.

[46] Gu, G.-F.; Zhou, W.-X. Emergence of long memory in stock volatility from a modified Mike-Farmer model. EPL 2009, 86, 48002.

[47] Cont, R.; Stoikov, S.; Talreja, R. A stochastic model for order book dynamics. Oper. Res. 2010, 58, 549-563.

[48] Yura, Y.; Takayasu, H.; Sornette, D.; et al. Financial Brownian particle in the layered order-book fluid and fluctuation-dissipation relations. Phys. Rev. Lett. 2014, 112, 098703.

[49] Mantegna, R.N. Hierarchical structure in financial markets.Eur. Phys. J. B 1999, 11, 193-197.

[50] Onnela, J.-P.; Chakraborti, A.; Kaski, K.; et al. Dynamics of market correlations: taxonomy and portfolio analysis. Phys. Rev. E 2003, 68, 056110.

[51] Tumminello, M.; Aste, T.; Di Matteo, T.; et al. A tool for filtering information in complex systems. Proc. Natl. Acad. Sci. USA 2005, 102, 10421-10426.

[52] David, J.; Di Matteo, T.; Jalan, S. Community detection in financial hypergraphs. Chaos, Solitons, and Fractals 2026, 212, 118953.

[53] Laloux, L.; Cizeau, P.; Bouchaud, J.-P.; et al. Noise dressing of financial correlation matrices. Phys. Rev. Lett. 1999, 83, 1467-1470.

[54] Plerou, V.; Gopikrishnan, P.; Rosenow, B.; et al. Universal and nonuniversal properties of cross correlations in financial time series. Phys. Rev. Lett. 1999, 83, 1471-1474.

[55] Raddant, M.; Di Matteo, T. A look at financial dependencies by means of econophysics and financial economics.J. Econ. Interact. Coord. 2023, 18, 701-734.

[56] Takayasu, H.; Miura, H.; Hirabayashi, T.; et al. Statistical properties of deterministic threshold elements - the case of market price. Physica A 1992, 184, 127-134.

[57] Palmer, R.G.; Arthur, W.B.; Holland, J.H.; et al. Artificial economic life: a simple model of a stockmarket. Physica D 1994, 75, 264-274.

[58] Bak, P.; Paczuski, M.; Shubik, M. Price variations in a stock market with many agents. Physica A 1997, 246, 430-453.

[59] Challet, D.; Marsili, M.; Zhang, Y.-C. Minority Games: Interacting Agents in Financial Markets. Oxford University Press, Oxford, 2005.

[60] Lux, T.; Marchesi, M. Scaling and criticality in a stochastic multi-agent model of a financial market. Nature 1999, 397, 498-500.

[61] Alfi, V.; Cristelli, M.; Pietronero, L.; et al. Minimal agent-based model for financial markets I: origin and self-organization of stylized facts. Eur. Phys. J. B 2009, 67, 385-397.

[62] Kanazawa, K.; Sueshige, T.; Takayasu, H.; et al. Derivation of the Boltzmann equation for financial Brownian motion: Direct observation of the collective motion of high-frequency traders. Phys. Rev. Lett. 2018, 120, 138301.

[63] Axtell, R.L.; Farmer, J.D. Agent-based modeling in economics and finance: Past, present, and future. J. Econ. Lit. 2025, 63, 197-287.

[64] He, Z.; Song, Q.; Lian, J.; et al. Towards standardizing validation practices in agent-based modeling: A hierarchical ABM validation framework. ACM Trans. Model. Comput. Simul. 2026, 36, 4:1-4:25.

[65] Jackson, M.O. Networks in the understanding of economic behaviors. J. Econ. Perspect. 2014, 28, 3-22.

[66] Acemoglu, D.; Azar, P.D. Endogenous production networks. Econometrica 2020, 88, 33-82.

[67] Battiston, S.; Caldarelli, G.; May, R.M.; et al. The price of complexity in financial networks. Proc. Natl. Acad. Sci. USA 2016, 113, 10031-10036.

[68] Chakraborti, A.; Chakrabarti, B.K. Statistical mechanics of money: how saving propensity affects its distribution. Eur. Phys. J. B 2000, 17, 167-170.

[69] Drăgulescu, A.; Yakovenko, V.M. Statistical mechanics of money. Eur. Phys. J. B 2000, 17, 723-729.

[70] Hidalgo, C.A.; Hausmann, R. The building blocks of economic complexity. Proc. Natl. Acad. Sci. USA 2009, 106, 10570-10575.

[71] Tacchella, A.; Cristelli, M.; Caldarelli, G.; et al. Economic complexity: conceptual grounding of a new metrics for global competitiveness. J. Econ. Dyn. Control 2013, 37, 1683-1691.

[72] Battiston, S.; Puliga, M.; Kaushik, R.; et al. DebtRank: too central to fail? Financial networks, the FED and systemic risk. Sci. Rep. 2012, 2, 541.

[73] Diem, C.; Borsos, A.; Reisch, T.; et al. Quantifying firm-level economic systemic risk from nation-wide supply networks. Sci. Rep. 2022, 12, 7719.

[74] Klimek, P.; Hausmann, R.; Thurner, S. Empirical confirmation of creative destruction from world trade data. PLoS ONE 2012, 7, e38924.

[75] Iacopini, I.; Milojević, S.; Latora, V. Network dynamics of innovation processes. Phys. Rev. Lett. 2018, 120, 048301.

[76] Kauffman, S.A. Investigations. Oxford University Press, Oxford, 2000.

[77] Rizzo, L.; Lee, E.; Kertész, J. Adjacent possible innovation dynamics on local optima networks. arXiv 2026, arXiv:2605.01821.

[78] Kim, D.; Cerigo, D.B.; Jeong, H.; et al. Technological novelty profile and invention's future impact.EPJ Data Sci. 2016, 5, 8.

[79] André, J.-B.; Cozic, M.; de Monte, A.; et al. From Evolutionary Biology to Economics and Back. Springer, Cham, 2023.

[80] Watts, D.J.; Strogatz, S.H. Collective dynamics of ‘small-world’ networks. Nature 1998, 393, 440-442.

[81] McPherson, M.; Smith-Lovin, L.; Cook, J.M. Birds of a feather: homophily in social networks. Annu. Rev. Sociol. 2001, 27, 415-444.

[82] Barabási, A.-L.; Albert, R. Emergence of scaling in random networks. Science 1999, 286, 509-512.

[83] Newman, M.E.J. Networks, 2nd ed.; Oxford University Press, Oxford, 2018.

[84] Barabási, A.-L. Network Science. Cambridge University Press, Cambridge, 2016.

[85] Battiston, F.; Cencetti, G.; Iacopini, I.; et al. Networks beyond pairwise interactions: structure and dynamics. Phys. Rep. 2020, 874, 1-92.

[86] Holme, P.; Saramäki, J. Temporal networks. Phys. Rep. 2012, 519, 97-125.

[87] Aleta, A.; et al. Multilayer network science. J. Complex Netw. 2026, 14, cnag007.

[88] Díaz-Díaz, F.; et al. Signed networks: theory, methods, and applications. arXiv 2025, arXiv:2511.17247.

[89] Barrat, A.; Barthélemy, M.; Pastor-Satorras, R.; et al. The architecture of complex weighted networks. Proc. Natl. Acad. Sci. USA 2004, 101, 3747-3752.

[90] Wan, J.; Ichinose, G.; Small, M.; et al. Multilayer networks with higher-order interaction reveal the impact of collective behavior on epidemic dynamics. Chaos Solitons Fractals 2022, 164, 112735.

[91] Montanari, A.; Saberi, A. The spread of innovations in social networks. Proc. Natl. Acad. Sci. USA 2010, 107, 20196-20201.

[92] Peralta, A.F.; Ramaciotti, P.; Kertész, J.; et al. Multidimensional political polarization in online social networks. Phys. Rev. Res. 2024, 6, 013170.

[93] Ohme, J.; Araujo, T.; Boeschoten, L.; et al. Digital trace data collection for social media effects research: APIs, data donation, and (screen) tracking. Commun. Methods Meas. 2024, 18, 124-141.

[94] Onnela, J.-P.; Saramäki, J.; Hyvönen, J.; et al. Structure and tie strengths in mobile communication networks. Proc. Natl. Acad. Sci. USA 2007, 104, 7332-7336.

[95] Palla, G.; Barabási, A.-L.; Vicsek, T. Quantifying social group evolution. Nature 2007, 446, 664-667.

[96] Blondel, V.D.; Decuyper, A.; Krings, G. A survey of results on mobile phone datasets analysis. EPJ Data Sci. 2015, 4, 10.

[97] Newman, M.E.J. Network structure from rich but noisy data.Nat. Phys. 2018, 14, 542-545.

[98] Cimini, G.; Squartini, T.; Saracco, F.; et al. The statistical physics of real-world networks. Nat. Rev. Phys. 2019, 1, 58-71.

[99] Dunbar, R.I.M. Neocortex size as a constraint on group size in primates. J. Hum. Evol. 1992, 22, 469-493.

[100] Saramäki, J.; Leicht, E.A.; López, E.; et al. Persistence of social signatures in human communication. Proc. Natl. Acad. Sci. USA 2014, 111, 942-947.

[101] Zhou, W.-X.; Sornette, D.; Hill, R.A.; et al. Discrete hierarchical organization of social group sizes. Proc. R. Soc. B 2005, 272, 439-444.

[102] Murase, Y.; Török, J.; Jo, H.-H.; et al. Multilayer weighted social network model. Phys. Rev. E 2014, 90, 052810.

[103] Fortunato, S.; Hric, D. Community detection in networks: A user guide. Phys. Rep. 2016, 659, 1-44.

[104] Fortunato, S.; Barthélemy, M. Resolution limit in community detection. Proc. Natl. Acad. Sci. USA 2007, 104, 36-41.

[105] Peixoto, T.P. Descriptive vs. Inferential Community Detection in Networks: Pitfalls, Myths and Half-Truths. Cambridge University Press, Cambridge, 2023.

[106] Palla, G.; Derényi, I.; Farkas, I.; et al. Uncovering the overlapping community structure of complex networks in nature and society. Nature 2005, 435, 814-818.

[107] Newman, M.E.J.; Girvan, M. Finding and evaluating community structure in networks. Phys. Rev. E 2004, 69, 026113.

[108] Ahn, Y.-Y.; Bagrow, J.P.; Lehmann, S. Link communities reveal multiscale complexity in networks. Nature 2010, 466, 761-764.

[109] Kivelä, M.; Arenas, A.; Barthelémy, M.; et al. Multilayer networks. J. Complex Netw. 2014, 2, 203-271.

[110] Peixoto, T.P.; Peel, L.; Gross, T.; et al. Graphs are maximally expressive for higher-order interactions. arXiv 2026, arXiv:2602.16937.

[111] Masuda, N.; Lambiotte, R. A Guide to Temporal Networks. World Scientific, Singapore, 2020.

[112] de Meijere, G.; Karsai, M.; Iñiguez, G. Universal time scales linking topology and dynamics in temporal networks. arXiv 2026, arXiv:2606.21696.

[113] Kitsak, M.; Gallos, L.K.; Havlin, S.; et al. Identification of influential spreaders in complex networks. Nat. Phys. 2010, 6, 888-893.

[114] Masuda, N.; Porter, M.A.; Lambiotte, R. Random walks and diffusion on networks. Phys. Rep. 2017, 716-717, 1-58.

[115] Pastor-Satorras, R.; Castellano, C.; Van Mieghem, P.; et al. Epidemic processes in complex networks. Rev. Mod. Phys. 2015, 87, 925-979.

[116] Schlosser, F.; Maier, B.F.; Jack, O.; et al. COVID-19 lockdown induces disease-mitigating structural changes in mobility networks. Proc. Natl. Acad. Sci. USA 2020, 117, 32883-32890.

[117] Kraemer, M.U.G.; Yang, C.-H.; Gutierrez, B.; et al. The effect of human mobility and control measures on the COVID-19 epidemic in China. Science 2020, 368, 493-497.

[118] Chinazzi, M.; et al. The effect of travel restrictions on the spread of the 2019 novel coronavirus (COVID-19) outbreak. Science 2020, 368, 395-400.

[119] Aleta, A.; et al. Modelling the impact of social distancing, testing, contact tracing and household quarantine on second-wave scenarios of the COVID-19 epidemic. Nat. Hum. Behav. 2020, 4, 964-971.

[120] Centola, D.; Macy, M. Complex contagions and the weakness of long ties. Am. J. Sociol. 2007, 113, 702-734.

[121] Andres, E.; Pastor-Satorras, R.; Starnini, M.; et al. Competition between simple and complex contagion on temporal networks. Phys. Rev. Res. 2025, 7, 043088.

[122] Lenti, J.; Mejova, Y.; Kalimeri, K.; et al. Global misinformation spillovers in the vaccination debate before and during the COVID-19 pandemic: Multilingual Twitter study. JMIR Infodemiology 2023, 3, e44714.

[123] Mønsted, B.; Lehmann, S. Characterizing polarization in online vaccine discourse-A large-scale study. PLoS ONE 2022, 17, e0263746.

[124] Caldarelli, G.; Artime, O.; Fischetti, G.; et al. The physics of news, rumors, and opinions. Phys. Rep. 2026, 1186, 1-75.

[125] Vosoughi, S.; Roy, D.; Aral, S. The spread of true and false news online. Science 2018, 359, 1146-1151.

[126] Lazer, D.M.J.; et al. The science of fake news. Science 2018, 359, 1094-1096.

[127] Zhao, Z.; Zhao, J.; Sano, Y.; et al. Fake news propagates differently from real news even at early stages of spreading. EPJ Data Sci. 2020, 9, 7.

[128] Oda, H.; Makovi, K.; Yasseri, T.; et al. A field experiment of social influence and behavioral contagion with bots on Reddit. arXiv 2026, arXiv:2607.00854.

[129] Lü, L.; Medo, M.; Yeung, C.H.; et al. Recommender systems. Phys. Rep. 2012, 519, 1-49.

[130] Pan, L.; Wang, C.-Y.; Zhou, F.; et al. Complexity of social media in the era of generative AI. Natl. Sci. Rev. 2025, 12, nwae323.

[131] Castellano, C.; Fortunato, S.; Loreto, V. Statistical physics of social dynamics. Rev. Mod. Phys. 2009, 81, 591-646.

[132] DeGroot, M.H. Reaching a consensus. J. Am. Stat. Assoc. 1974, 69, 118-121.

[133] Friedkin, N.E.; Johnsen, E.C. Social influence and opinions.J. Math. Sociol. 1990, 15, 193-206.

[134] Deffuant, G.; Neau, D.; Amblard, F.; et al. Mixing beliefs among interacting agents. Adv. Complex Syst. 2000, 3, 87-98.

[135] Hegselmann, R.; Krause, U. Opinion dynamics and bounded confidence models, analysis, and simulation. J. Artif. Soc. Soc. Simul. 2002, 5(3), 2.

[136] Peralta, A.F.; Kertész, J.; Iñiguez, G. Opinion dynamics in social networks: From models to data. In Handbook of Computational Social Science; Yasseri, T., Ed.; Edward Elgar, Cheltenham, 2025.

[137] Fernández-Gracia, J.; Suchecki, K.; Ramasco, J.J.; et al. Is the voter model a model for voters? Phys. Rev. Lett. 2014, 112, 158701.

[138] Levin, S.A.; Milner, H.V.; Perrings, C. The dynamics of political polarization. Proc. Natl. Acad. Sci. USA 2021, 118, e2116950118.

[139] Górski, P.J.; Vaccario, G.; Hołyst, J.A.; et al. More connections, higher polarization: The role of weak ties and status-driven dynamics. Proc. Natl. Acad. Sci. USA 2026, 123, e2614422123.

[140] Galam, S. Minority opinion spreading in random geometry.Eur. Phys. J. B 2002, 25, 403-406.

[141] Cinelli, M.; De Francisci Morales, G.; Galeazzi, A.; et al. The echo chamber effect on social media. Proc. Natl. Acad. Sci. USA 2021, 118, e2023301118.

[142] Blex, C.; Yasseri, T. Positive algorithmic bias cannot stop fragmentation in homophilic networks. J. Math. Sociol. 2022, 46, 80-97.

[143] Lorenz-Spreen, P.; Mønsted, B.M.; Hövel, P.; et al. Accelerating dynamics of collective attention. Nat. Commun. 2019, 10, 1759.

[144] Karsai, M.; Jo, H.-H.; Kaski, K. Bursty Human Dynamics. Springer, Cham, 2018.

[145] Pappalardo, L.; et al. A survey on the impacts of recommender systems on users, items, and human-AI ecosystems. arXiv 2024, arXiv:2407.01630.

[146] Piaggesi, S.; Giurgola, S.; Karsai, M.; et al. Mapping urban socioeconomic inequalities in developing countries through Facebook advertising data. Front. Big Data 2022, 5, 1006352.

[147] Espín-Noboa, L.; Kertész, J.; Karsai, M. Interpreting wealth distribution via poverty map inference using multimodal data. In Proceedings of the ACM Web Conference 2023; pp. 4029-4040.

[148] Manna, A.; Koltai, J.; Karsai, M. Importance of social inequalities to contact patterns, vaccine uptake, and epidemic dynamics. Nat. Commun. 2024, 15, 4137.

[149] Karimi, F.; Oliveira, M.; Strohmaier, M. Minorities in networks and algorithms. In Handbook of Computational Social Science; Yasseri, T., Ed.; Edward Elgar, Cheltenham, 2025.

[150] Schelling, T.C. Dynamic models of segregation. J. Math. Sociol. 1971, 1, 143-186.

[151] Vega-Redondo, F. Economics and the Theory of Games. Cambridge University Press, Cambridge, 2010.

[152] Lucas, A. Nonequilibrium phase transitions in competitive markets caused by network effects. Proc. Natl. Acad. Sci. USA 2022, 119, e2206702119.

[153] Szabó, G.; Fáth, G. Evolutionary games on graphs. Phys. Rep. 2007, 446, 97-216.

[154] Nowak, M.A. Five rules for the evolution of cooperation. Science 2006, 314, 1560-1563.

[155] Perc, M.; Gómez-Gardeñes, J.; Szolnoki, A.; et al. Evolutionary dynamics of group interactions on structured populations: a review. J. R. Soc. Interface 2013, 10, 20120997.

[156] Gracia-Lázaro, C.; Ferrer, A.; Ruiz, G.; et al. Heterogeneous networks do not promote cooperation when humans play a Prisoner's Dilemma. Proc. Natl. Acad. Sci. USA 2012, 109, 12922-12926.

[157] Guido, A.; Andrighetto, G.; Sánchez, A.; et al. Disentangling material, social, and cognitive determinants of human behavior and beliefs. Humanit. Soc. Sci. Commun. 2023, 10, 236.

[158] Yasseri, T.; Kertész, J. Value production in a collaborative environment. J. Stat. Phys. 2013, 151, 414-439.

[159] Vedres, B.; Vasarhelyi, O. Gendered behavior as a disadvantage in open-source software development. EPJ Data Sci. 2019, 8, 25.

[160] Johnson, N.F.; Carran, S.; Botner, J.; et al. Pattern in escalations in insurgent and terrorist activity. Science 2011, 333, 81-84.

[161] Galam, S.; Javarone, M.A. Modeling radicalization phenomena in heterogeneous populations. PLoS ONE 2016, 11, e0155407.

[162] Wang, Y.; Bu, F. Modeling radicalization of terrorism under the influence of multiple ideologies. AIMS Math. 2022, 7, 4833-4850.

[163] González, M.C.; Hidalgo, C.A.; Barabási, A.-L. Understanding individual human mobility patterns. Nature 2008, 453, 779-782.

[164] Song, C.; Koren, T.; Wang, P.; et al. Modelling the scaling properties of human mobility.Nat. Phys. 2010, 6, 818-823.

[165] Schneider, C.M.; Belik, V.; Couronné, T.; et al. Unravelling daily human mobility motifs. J. R. Soc. Interface 2013, 10, 20130246.

[166] Simini, F.; González, M.C.; Maritan, A.; et al. A universal model for mobility and migration patterns. Nature 2012, 484, 96-100.

[167] Lu, X.; Feng, J.; Lai, S.; et al. Human mobility in epidemic modeling. Phys. Rep. 2026, 1157, 1-45.

[168] Louail, T.; Lenormand, M.; Cantu Ros, O.G.; et al. From mobile phone data to the spatial structure of cities. Sci. Rep. 2014, 4, 5276.

[169] Rybski, D.; González, M.C. Cities as complex systems-Collection overview. PLoS ONE 2022, 17, e0262964.

[170] Bettencourt, L.M.A.; Lobo, J.; Helbing, D.; et al. Growth, innovation, scaling, and the pace of life in cities. Proc. Natl. Acad. Sci. USA 2007, 104, 7301-7306.

[171] West, G. Scale: The Universal Laws of Life, Growth, and Death in Organisms, Cities, and Companies. Penguin Books, New York, 2018.

[172] Tóth, G.; Wachs, J.; Di Clemente, R.; et al. Inequality is rising where social network segregation interacts with urban topology.Nat. Commun. 2021, 12, 1143.

[173] Caldarelli, G.; Arcaute, E.; Barthelemy, M.; et al. The role of complexity for digital twins of cities. Nat. Comput. Sci. 2023, 3, 374-381.

[174] Rodríguez, J.P.; Aleta, A.; Moreno, Y. Multilayer networks describing urban interactions for building the digital twins of five cities in Spain. Sci. Data 2025, 12, 1227.

[175] Wang, C.-Y.; Huang, D.F.; Zhu, X.Y.; et al. Cyber-physical-social system and social governance in the digital-intelligent era. Sci. Sin. Inf. 2026, 56, 1-25.

[176] Barthelemy, M. The Structure and Dynamics of Cities. Cambridge University Press, Cambridge, 2016.

[177] Caldarelli, G.; Chiesi, L.; Chirici, G.; et al. Lessons from complex networks to smart cities. Nat. Cities 2025, 2, 127-134.

[178] Lighthill, M.J.; Whitham, G.B. On kinematic waves. II. A theory of traffic flow on long crowded roads. Proc. R. Soc. Lond. A 1955, 229, 317-345.

[179] Newell, G.F. Nonlinear effects in the dynamics of car following. Oper. Res. 1961, 9, 209-229.

[180] Nagel, K.; Schreckenberg, M. A cellular automaton model for freeway traffic. J. Phys. I France 1992, 2, 2221-2229.

[181] Helbing, D. Traffic and related self-driven many-particle systems. Rev. Mod. Phys. 2001, 73, 1067-1141.

[182] Nagatani, T. The physics of traffic jams. Rep. Prog. Phys. 2002, 65, 1331-1386.

[183] Ivanova, E.M. Mathematical models for traffic flow optimization.Am. J. Eng. Math. 2020, 1, 1-7.

[184] von Ferber, C.; Holovatch, T.; Holovatch, Y.; et al. Public transport networks: Empirical analysis and modeling. Eur. Phys. J. B 2009, 68, 261-275.

[185] Aleta, A.; Meloni, S.; Moreno, Y. A multilayer perspective for the analysis of urban transportation systems. Sci. Rep. 2017, 7, 44359.

[186] Li, D.; Fu, B.; Wang, Y.; et al. Percolation transition in dynamical traffic network with evolving critical bottlenecks. Proc. Natl. Acad. Sci. USA 2015, 112, 669-672.

[187] Monechi, B.; Gravino, P.; Di Clemente, R.; et al. Complex delay dynamics on railway networks: From universal laws to realistic modelling. EPJ Data Sci. 2018, 7, 35.

[188] Zheng, Y.Y.; Takayasu, H.; Takayasu, M. Do railway commuters exhibit consistent route choice 'rationality' across different contexts and time? Evidence from Tokyo metropolitan commutes. J. Soc. Phys. 2026, 1, 6.

[189] Helbing, D.; Farkas, I.; Vicsek, T. Simulating dynamical features of escape panic. Nature 2000, 407, 487-490.

[190] Zhong, Z.; Takayasu, H.; Takayasu, M. Renormalization of human mobility based on a revised electric circuit model and a new gravity relation. Phys. Rev. Res. 2025, 7, 013235.

[191] Altmann, E.G.; Gerlach, M. Statistical laws in linguistics. In Creativity and Universality in Language; Esposti, M.D.; Altmann, E.G.; Pachet, F., Eds.; Springer, Cham, 2016; pp. 7-26.

[192] Abrams, D.M.; Strogatz, S.H. Modelling the dynamics of language death. Nature 2003, 424, 900.

[193] Battiston, F.; Nicosia, V.; Latora, V.; et al. Layered social influence promotes multiculturality in the Axelrod model. Sci. Rep. 2017, 7, 1809.

[194] Gessey-Jones, T.; Connaughton, C.; Dunbar, R.; et al. Narrative structure of A Song of Ice and Fire creates a fictional world with realistic measures of social complexity. Proc. Natl. Acad. Sci. USA 2020, 117, 28582-28588.

[195] Mac Carron, P., Kenna, R. Universal properties of mythological networks. EPL 2012, 99, 28002.

[196] Liu, X.F.; Tse, C.K.; Small, M. Complex network structure of musical compositions: Algorithmic generation of appealing music. Physica A 2010, 389, 126-132.

[197] McFee, B. Quantifying regularity in music structure analysis. In Proceedings of the 26th International Society for Music Information Retrieval Conference, Daejeon, Republic of Korea, 2025.

[198] Kim, D.; Son, S.-W.; Jeong, H. Large-scale quantitative analysis of painting arts. Sci. Rep. 2014, 4, 7370.

[199] Barabási, A.-L. The Formula: The Universal Laws of Success. Little, Brown, Boston, 2018.

[200] Huang, Z.-F.; Stauffer, D. Stochastic Penna model for biological aging. Theor. Biosci. 2001, 120, 21-27.

[201] Monsivais, D.; Bhattacharya, K.; Ghosh, A.; et al. Seasonal and geographical impact on human resting periods. Sci. Rep. 2017, 7, 10717.

[202] Strulik, H. A mass phenomenon: The social evolution of obesity. J. Health Econ. 2014, 33, 113-125.

[203] Cintia, P.; Pappalardo, L. Quantifying the relation between performance and success in soccer.Adv. Complex Syst. 2018, 21, 1750014.

[204] Asztalos, B.; Balázs, B.; Palla, G.; et al. Hierarchy and ranking in fencing and tennis. Sci. Rep. 2026, 16, 24877.

[205] Vicsek, T.; Czirók, A.; Ben-Jacob, E.; et al. Novel type of phase transition in a system of self-driven particles. Phys. Rev. Lett. 1995, 75, 1226-1229.

[206] Romanczuk, P., Daniels, B.C. Phase transitions and criticality in the collective behavior of animals - self-organization and biological function. In: Order, Disorder, and Criticality. Advanced Problems of Phase Transition Theory. Yu. Holovatch (editor). vol. 7 World Scientific, Singapore, 2023, 179-208. arXiv:2211.03879.

[207] Merton, R.K. The Matthew effect in science. Science 1968, 159, 56-63.

[208] Wang, D.; Barabási, A.-L. The Science of Science. Cambridge University Press, Cambridge, 2021.

[209] Newman, M.E.J. The structure of scientific collaboration networks. Proc. Natl. Acad. Sci. USA 2001, 98, 404-409.

[210] Barabási, A.-L.; Jeong, H.; Néda, Z.; et al. Evolution of the social network of scientific collaborations. Physica A 2002, 311, 590-614.

[211] Price, D.J. de S. A general theory of bibliometric and other cumulative advantage processes. J. Am. Soc. Inf. Sci. 1976, 27, 292-306.

[212] Wuchty, S.; Jones, B.F.; Uzzi, B. The increasing dominance of teams in production of knowledge. Science 2007, 316, 1036-1039.

[213] Uzzi, B.; Mukherjee, S.; Stringer, M.; et al. Atypical combinations and scientific impact. Science 2013, 342, 468-472.

[214] Liu, L.; Wang, Y.; Sinatra, R.; et al. Hot streaks in artistic, cultural, and scientific careers. Nature 2018, 559, 396-399.

[215] Jalan, S.; Sarkar, C.; Madhusudanan, A.; et al. Uncovering randomness and success in society. PLoS ONE 2014, 9, e88249.

[216] Radicchi, F.; Fortunato, S.; Castellano, C. Universality of citation distributions: Toward an objective measure of scientific impact. Proc. Natl. Acad. Sci. USA 2008, 105, 17268-17272.

[217] Majka, A.; El-Mhamdi, E.M. The strong, weak and benign Goodhart's law. arXiv 2025, arXiv:2505.23445.

[218] Fire, M.; Guestrin, C. Over-optimization of academic publishing metrics: Observing Goodhart's Law in action. GigaScience 2019, 8, giz053.

[219] Zheng, J. A review and methodological reflection on knowledge organization studies. In Proceedings of ICELAIC 2021; Atlantis Press, Paris, 2022.

[220] Bagnoli, F.; de Bonfioli Cavalcabo, G. A simple model of knowledge percolation. In Cellular Automata; Chopard, B.; Bandini, S.; Dennunzio, A.; Arabi Haddad, M., Eds.; Lecture Notes in Computer Science, Vol. 13402; Springer, Cham, 2022.

[221] Börner, K.; Chen, C.; Boyack, K.W. Visualizing knowledge domains. Annu. Rev. Inf. Sci. Technol. 2003, 37, 179-255.

[222] Piccardi, T.; West, R. Navigating knowledge: Patterns and insights from Wikipedia consumption. In Handbook of Computational Social Science; Yasseri, T., Ed.; Edward Elgar, Cheltenham, 2025.

[223] Zhu, M.; Yasseri, T.; Kertész, J. Individual differences in knowledge network navigation. Sci. Rep. 2024, 14, 8331.

[224] Wikipedia contributors. Wikipedia: Wikipedia games. Available online: https://en.wikipedia.org/wiki/Wikipedia:Wikipedia_games (accessed 25 August 2026).

[225] Tsvetkova, M.; Yasseri, T.; Pescetelli, N.; et al. A new sociology of humans and machines. Nat. Hum. Behav. 2024, 8, 1864-1876.

[226] Pedreschi, D.; Pappalardo, L.; Ferragina, E.; et al. Human-AI coevolution. Artif. Intell. 2025, 339, 104244.

[227] Kordzadeh, N.; Ghasemaghaei, M. Algorithmic bias: Review, synthesis, and future research directions. Eur. J. Inf. Syst. 2022, 31, 388-409.

[228] Sîrbu, A.; Pedreschi, D.; Giannotti, F.; et al. Algorithmic bias amplifies opinion fragmentation and polarization: A bounded confidence model. PLoS ONE 2019, 14, e0213246.

[229] Shumailov, I.; Shumaylov, Z.; Zhao, Y.; et al. AI models collapse when trained on recursively generated data. Nature 2024, 631, 755-759.

[230] Brinkmann, L.; Baumann, F.; Bonnefon, J.-F.; et al.Machine culture. Nat. Hum. Behav. 2023, 7, 1855-1868.

[231] Cui, H.; Yasseri, T. AI-enhanced collective intelligence. Patterns 2024, 5, 101074.

[232] Burton, J.W.; Lopez-Lopez, E.; Hecht, M.; et al. How large language models can reshape collective intelligence. Nat. Hum. Behav. 2024, 8, 1643-1655.

[233] Törnberg, P.; Uitermark, J. Social physics in the age of generative AI: How, exactly, is society complex? Sociologica 2026, 20, 141-150.

[234] Han, T.A.; Leibo, J.Z.; Lenaerts, T.; et al. Social Physics in the age of artificial intelligence. arXiv 2026, arXiv:2603.16900.

[235] Bazazi, S.; Karpus, J.; Yasseri, T. AI's assigned gender affects human-AI cooperation. iScience 2025, 28, 113905.

[236] Vaccaro, M.; Almaatouq, A.; Malone, T.W. When combinations of humans and AI are useful: A systematic review and meta-analysis.Nat. Hum. Behav. 2024, 8, 2293-2303.

[237] Samadi, M.A.; Martins de Bastos, P.M.; Choi, J.; et al. TRAIL: A platform for configurable human-AI teaming experiments. arXiv 2026, arXiv:2607.12180.

[238] Lin, Z.; Shen, S.; Cheng, Z.; et al. Carbon and silicon, coexist or compete? A survey on human-AI interactions in agent-based modeling and simulation. arXiv 2025, arXiv:2502.18145.

[239] Su, C.; Wang, C.-Y.; Shi, W.-F.; et al. Generative sociology: Leveraging large language models to comprehend social complexity. Natl. Sci. Rev. 2026, nwag513.

[240] Li, A.; Chen, H.; Namkoong, H.; et al.LLM generated persona is a promise with a catch. In Advances in Neural Information Processing Systems (NeurIPS), 2025.

[241] Minder, J.; et al. Synthetic persona pretraining: Alignment from token zero. arXiv 2026, arXiv:2608.13482.

[242] Ge, T.; Chan, X.; Wang, X.; et al. Scaling synthetic data creation with 1,000,000,000 personas. arXiv 2024, arXiv:2406.20094.

[243] Ashery, A.F.; Aiello, L.M.; Baronchelli, A. Emergent social conventions and collective bias in LLM populations. Sci. Adv. 2025, 11, eadu9368.

[244] Fadaei, F.; Moran, J.C.; Yasseri, T. Gender dynamics and homophily in a social network of LLM agents. Philos. Trans. R. Soc. B 2026, forthcoming; arXiv:2602.02606.

[245] Akata, E.; Schulz, L.; Coda-Forno, J.; et al. Playing repeated games with large language models. Nat. Hum. Behav. 2025, 9, 1380-1390.

[246] Rossi, L.; Harrison, K.; Shklovski, I. The problems of LLM-generated data in social science research. Sociologica 2024, 18, 145-168.

[247] Lazer, D.; Kennedy, R.; King, G.; et al. The parable of Google Flu: Traps in big data analysis. Science 2014, 343, 1203-1205.

[248] Blank, G. The digital divide among Twitter users and its implications for social research. Soc. Sci. Comput. Rev. 2017, 35, 679-697.

[249] Leifeld, P.; Martínez, S.; Brandenberger, L. Disentangling social contagion from prior similarity in time-ordered behavior sequences. J. R. Stat. Soc. Ser. A 2026, qnag069.

[250] Sasvári, P.; Lendvai, G.F. The overrepresentation of the United States in the field of legal studies. J. Informetr. 2025, 19, 101680.

[251] Gallegati, M.; Keen, S.; Lux, T.; et al. Worrying trends in econophysics. Physica A 2006, 370, 1-6.

[252] Simkin, M.V.; Roychowdhury, V.P. Re-inventing Willis. Physics Reports 2011, 502, 1-35.